\documentclass[aps,prb,twocolumn,groupedaddress,floatfix,letterpaper]{revtex4}

\usepackage{epsfig}
\usepackage{graphicx}
\usepackage{float}
\begin{document}

\title{Spatial imaging of modifications to fluorescence lifetime and intensity by individual Ag nanoparticles}

\author{T.\ Ritman-Meer}
\author{N.\ I.\ Cade}
\email{nicholas.cade@kcl.ac.uk}
\author{D.\ Richards}
\affiliation{Department of Physics, King's College London, Strand, London WC2R 2LS, UK}


\begin{abstract}
Highly ordered periodic arrays of silver nanoparticles have been fabricated which exhibit surface plasmon resonances in the visible spectrum. We
demonstrate the ability of these structures to alter the fluorescence properties of vicinal dye molecules by providing an additional radiative
decay channel. Using fluorescence lifetime imaging microscopy (FLIM), we have created high resolution spatial maps of the molecular lifetime
components; these show an order of magnitude increase in decay rate from a localized volume around the nanoparticles, resulting in a
commensurate enhancement in the fluorescence emission intensity.
\end{abstract}

\maketitle

Fluorescence spectroscopy and lifetime measurements are powerful tools in the study of biological samples;\cite{festy07} however, these
techniques suffer from limited sensitivity at low fluorophore concentrations typically employed. The presence of a metallic surface can
dramatically alter the emission properties of a locally situated fluorophore, resulting in enhanced fluorescence emission and greater
photostability by reducing the excited state lifetime.\cite{barnes1998,lakowicz2002} These plasmon-induced fluorescence modifications are
currently being investigated for a wide variety of nanostructured systems, such as metal-island films,\cite{gheddes2007} lithographically
patterned nanoparticle arrays,\cite{gerber07} and individual gold nanospheres.\cite{novotny2006, seelig07}  The strong dependence of these
effects on fluorophore-metal separation has been used to increase both lateral and longitudinal resolution in confocal microscopy, with
important consequences for biological imaging.\cite{alschinger03}

Here, we report the results of high spatial resolution time-resolved measurements on individual self-assembled Ag nanoparticles. Using
fluorescence lifetime imaging microscopy (FLIM), we have observed localized enhancements in the emission intensity and additional fluorescence
lifetime components from vicinal fluorophores. By correlating the initial emission intensity and lifetime resulting from these modified decay
channels, we attribute these enhancements to a greater photon recycling rate due to coupling with surface plasmons.

\begin{figure}[tb]
\begin{center}
\epsfig{file=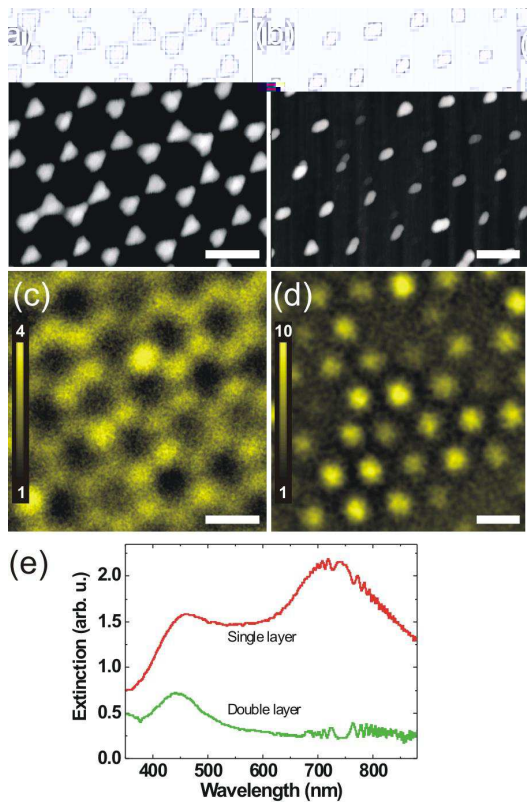,width=6.9cm}\end{center} \caption{(Color online) (a) \& (b) AFM images of Ag nanoparticles formed from single and double
layers of latex spheres, respectively. Scale bars = 500 nm and the height contrast is 25 nm for both.  (c) \& (d) Confocal CW fluorescence
intensity maps of R6G deposited on nanoparticle regions similar to (a) and (b), respectively. (e) Ensemble extinction spectra from single  and
double layer Ag nanoparticle arrays, showing dipole and quadrupole plasmon resonances. The high-frequency noise is from interference in the
glass slide.} \label{fluorescence}
\end{figure}

Aqueous 500 nm diameter latex spheres were drop-cast onto a piranha treated glass slide, resulting in the formation of a monolayer / bilayer
close-packed lattice as the water evaporated.\cite{vanduyne2001} A 0.5 nm layer of chromium was then deposited onto the slide by thermal
evaporation in a vacuum chamber; this increases the adhesion of silver which was subsequently deposited to a thickness of 25 nm. The latex
spheres were removed by sonication in chloroform, leaving a periodic array of Ag nanoparticles over a typical area of 20 mm$^2$. Figure
\ref{fluorescence}(a) shows an atomic force microscopy (AFM) image of a region of nanoparticles formed from a single layer of latex spheres;
individual particles have a base length of $\sim$150 nm and height of 25 nm. In some regions, different structures form due to Ag deposition on
double layers of spheres,\cite{vanduyne2001} as shown in Fig.\ \ref{fluorescence}(b).

Rhodamine 6G (R6G) dye was deposited onto the nanoparticles by vacuum sublimation to produce a uniform coverage of sub-monolayer thickness. The
sample was prepared for high resolution optical measurements by applying a thin layer of index-matching polymer to a glass slide and affixing
this on top of the nanoparticle array. During this procedure the R6G was incorporated into the polymer layer, creating a homogeneous
distribution of fluorophores across the sample with a thickness of several microns.

Fluorescence lifetime and intensity images were obtained using a scanning confocal epifluorescence microscope  (Leica TC SP2, 100x oil
objective, 1.4 NA) with a synchronous time correlated single photon counting module (Becker and Hickl SPC-830). Excitation was with a 488 nm
continuous-wave (CW) Ar$^{+}$ laser (intensity) or 467 nm 20 MHz pulsed diode laser (lifetime), at a power far below fluorescence saturation in
both cases.

Figures \ref{fluorescence}(c) and (d) show CW fluorescence intensity maps of R6G on single and double layer nanoparticles, respectively. These
maps show highly localized fluorescence enhancement from molecules close to nanoparticles. Spectral analysis has verified that the emission
originates from R6G and is not caused by photoluminescence from the silver nanoparticles or scattered laser light. The R6G fluorescence
enhancement measured from the nanoparticles, relative to that from the glass, is approximately fourfold and tenfold for (c) and (d),
respectively. The actual enhancement in the vicinity of the nanoparticles will be much larger than that measured: enhancement is only expected
to occur over a very small range of metal-fluorophore distances ($\sim$20 nm),\cite{novotny2006} hence there is a significant unmodified
fluorescence background from the other fluorophores in the excited confocal volume. This is discussed further below.

Extinction spectroscopy (Perkin-Elmer Lambda 800) was used to determine the local surface plasmon resonances (LSPR) of the Ag nanoparticles.
Figure \ref{fluorescence}(e) shows ensemble extinction spectra from single and double layer regions. The single layer structures have two strong
absorption resonances corresponding to dipole (720 nm) and quadrupole (460 nm) LSPR modes,\cite{schatz2003,nelayah2007} whereas the double layer
structures have only one peak at 440 nm due to their ellipsoidal geometry.\cite{vanduyne2001}

There are two main causes of fluorescence \emph{enhancement} from molecules close to metal nanoparticles: when optically excited at its plasmon
resonance, a nanoparticle can dramatically enhance the local electromagnetic field intensity, resulting in an increase in a molecule's
excitation rate.\cite{schatz2003,nelayah2007} The excited-state molecular dipole can also couple with surface plasmon electrons in the metal
creating an additional radiative decay channel.\cite{barnes1998,lakowicz2005,gheddes2007}

For CW measurements it is not possible to separate the contributions to the fluorescence enhancement arising from the modified excitation and
decay channels; however, this is not the case with time-resolved measurements. The quantum efficiency $Q$ gives the probability of an excited
state molecule decaying to a lower state by photon emission. In free-space $Q_{0}=\Gamma \tau_{0}$, where $\Gamma$ is the radiative emission
rate, $\tau_{0} = (\Gamma + k_{\mathrm{nr}})^{-1}$ is the total lifetime of the molecule, and $k_{\mathrm{nr}}$  is the sum of the non-radiative
decay rates. The presence of a metal creates an additional decay channel $\Gamma_{\mathrm{m}}$ for an excited molecule, so the total radiative
decay rate becomes $\Gamma^{\prime}=\Gamma+\Gamma_{\mathrm{m}}$. The modified fluorescence quantum yield $Q_{\mathrm{m}}$ and lifetime
$\tau_{\mathrm{m}}$ are then related by $Q_{\mathrm{m}}=\Gamma^{\prime}\tau_{\mathrm{m}}$, and $\tau_{\mathrm{m}} = (\Gamma^{\prime}+
k_{\mathrm{nr}}^{\prime})^{-1}$.

\begin{figure}[tb]
\begin{center}
\epsfig{file=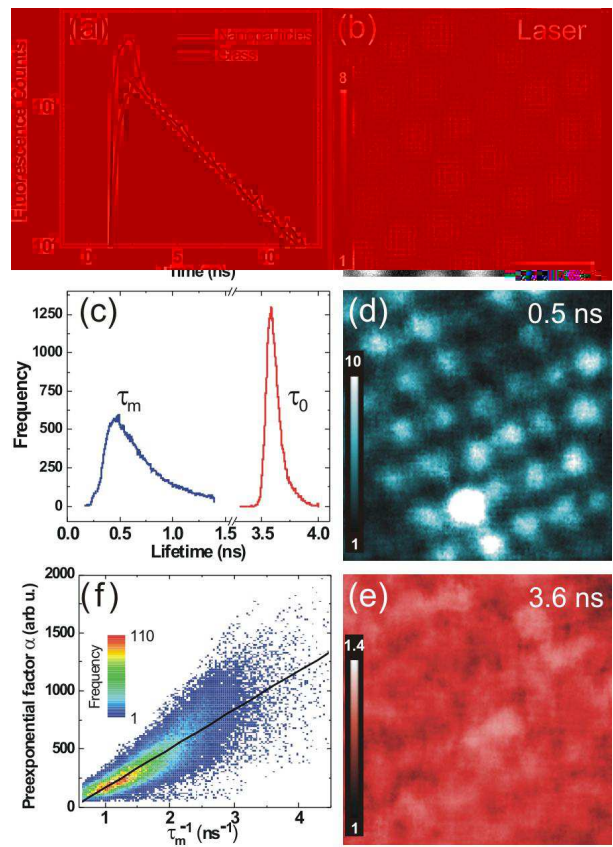,width=8cm}\end{center} \caption{(Color online) Fluorescence lifetime analysis for a region of double-layer nanoparticles:
(a) Raw decay transients from R6G on glass and nanoparticles; the straight line is a linear fit to the glass data. (b) Reflected laser intensity
from a small scan area; scale bar is 1 $\mu$m. (c) Histogram of lifetime components ($\tau_{\mathrm{m}}$, $\tau_{0}$) obtained from
biexponential pixel fits of the scan area. (d) Spatial map of the normalized preexponential intensity for the 0.5 ns $\tau_{\mathrm{m}}$
component. (e) as (d) for the 3.6 ns $\tau_{0}$ component, showing homogeneity across the region. Note the difference in the relative intensity
scales. (f) Pixel-by-pixel correlation between spatial maps of $\tau_{\mathrm{m}}$ and associated $\alpha$ values, for a $5\times5$ $\mu$m area
of nanoparticles. The color bar gives the binned frequency, and the line is a linear best fit.} \label{lifetime}
\end{figure}

The effects of Ag nanoparticles on the R6G fluorescence lifetime were investigated for a region of double-layer structures. After excitation by
a short laser pulse, the time evolution of the fluorescence intensity of a free-space single molecule is $I(t)= \alpha e^{-t /\tau_{0}}$. Figure
\ref{lifetime}(a) shows spatially integrated raw decay transients from large areas of glass and nanoparticles. In both cases there is an
identical long-lived component originating from unmodified fluorophores in the polymer layer; this has a monoexponential decay $>$3 ns and
implies that there are no significant concentration-induced non-radiative channels.\cite{leitner85}

Figure \ref{lifetime}(b) shows the reflected laser intensity map for a scan area similar to that of Figs.\ \ref{fluorescence}(b) and (d). A FLIM
map of this region was acquired for a 30 min integration period. Bins of $9\times9$ pixels ($20\times20$ nm) were used to give decay transients
with total preexponential counts of $\sim$$10^3$. Biexponential fits were then applied to give spatially resolved maps of constituent lifetimes
$\tau_{i}$ and their corresponding intensities $\alpha_{i}$.\cite{note1} A histogram of the extracted lifetime components is shown in Fig.\
\ref{lifetime}(c): this comprises a sharply peaked component $\tau_{0}$ from the unperturbed fluorophores, and a broader fast component
$\tau_{\mathrm{m}}$ arising from fluorophores with modified decay rates. Figures \ref{lifetime}(d) and (e) show maps of the normalized
preexponential intensities for the two modal lifetime components; the 0.5 ns component is localized (resolution limited) around the
nanoparticles, whereas the 3.6 ns component has a uniform intensity (standard deviation $< 2 \%$) consistent with the homogeneous coverage of
fluorophores.

For this sample, the R6G (in polymer) layer thickness is comparable to the axial confocal excitation depth; hence, fluorophores at different
distances from the nanoparticles will contribute differently to the total signal measured. Molecules in very close proximity to a metal show
strong fluorescence quenching due to dominant non-radiative energy transfer.\cite{dulkeith02} We have taken measurements on samples without the
additional polymer layer; these show almost complete quenching of the R6G fluorescence over the Ag nanoparticles, and an instrumental-response
limited lifetime. This strong distance-dependent weighting means that, for the current sample, the \emph{enhanced} fluorescence with a lifetime
$\tau_{\mathrm{m}}$ is from a thin shell of molecules around the nanoparticles where the competing quenching and enhancement mechanisms result
in a maximum net emission intensity.\cite{huang,novotny2006}

An aqueous R6G molecule has $Q=0.95$; hence, we assume that fluorophores showing a maximally enhanced emission intensity have $Q_{\mathrm{m}}
\approx 1$, so $\Gamma^{\prime} \approx \tau_{\mathrm{m}}^{-1}$. The preexponential factor $\alpha$ depends on several fixed experimental
parameters and also directly depends on the radiative decay rate $\Gamma^{\prime}$.\cite{gerber07} Thus, we expect a direct correlation between
$\alpha$ and the measured decay rate. Figure \ref{lifetime}(f) shows a pixel-by-pixel correlation between the spatial maps of the
$\tau_{\mathrm{m}}$ lifetime and their associated $\alpha$ values, extracted from biexponential fits over a region of nanoparticles as described
above. The data have been binned and the resulting frequency at each point is shown by the color bar. Scatter plots of $\chi^2$ versus $\alpha$
and $\tau$ were created from all biexponential fits; these show no fitting related bias. The data in Fig.\ \ref{lifetime}(f) are well fit by a
linear regression, in agreement with the above discussion. A large number of fluorophores show an order of magnitude increase in decay rate
relative to the unmodified background ($\tau_{0}^{-1}=0.28$ ns$^{-1}$), with a commensurate enhancement in initial intensity, as shown in Fig.\
\ref{lifetime}(d). In CW measurements, the nanoparticle-induced enhancement in integrated intensity cannot be accounted for by an increase in
$Q$. Instead it originates primarily from a greatly increased photon recycling rate for fluorophores with a strong radiative coupling to surface
plasmons.

In conclusion, we have demonstrated highly localized modification of fluorescence intensity and lifetime on ordered arrays of Ag nanoparticles.
FLIM maps of R6G fluorescence show an additional plasmon-induced radiative decay channel for a thin shell of molecules around individual
nanoparticles. This results in an order of magnitude increase in ensemble photon recycling rate with a proportionate enhancement in fluorescence
intensity. The strong distance dependence of these effects suggests that metal enhanced fluorescence may offer a means of selectively mapping
the location of specific target molecules within a larger excitation volume, such as fluorophore-tagged proteins in cell membranes.

The authors would like to thank F. Festy and K. Suhling for helpful discussions. This work was supported by the EPSRC (UK).


\begin{thebibliography}{}
\bibitem{note1} An instrumental response of $\sim$60 ps has been included in all lifetime analyis.
\end{thebibliography}
\end{document}